# Picosecond Spin-Orbit Torque Induced Coherent Magnetization Switching in a Ferromagnet

Debanjan Polley[1,2], Akshay Pattabi[2,3], Ashwin Rastogi[2], Kaushalya Jhuria[1,4], Eva Diaz[4], Hanuman Singh[2], Aristide Lemaitre[5], Michel Hehn[4], Jon Gorchon[4] and Jeffrey Bokor[1,2]

[1] Lawrence Berkeley National Laboratory, 1 Cyclotron Road, Berkeley, California 94720, USA
[2] Department of Electrical Engineering and Computer Sciences, University of California, Berkeley, California 94720, USA
[3] Department of Engineering, University of San Francisco, California 94117, USA
[4] Université de Lorraine, CNRS, IJL, Nancy, France.
[5] Université Paris-Saclay, CNRS, Centre de Nanosciences et de Nanotechnologies, 91120, Palaiseau, France

## Abstract

Electrically controllable non-volatile magnetic memories show great potential for the replacement of conventional semiconductor-based memory technologies. Recently there is strong interest in the spin-orbit torque (SOT) mechanism for magnetization reversal due to the devices' increased lifetime and speed of operation when compared to other alternatives. However, recent SOT switching studies reveal an incubation delay in the $\sim ns$ range due to stochasticity in the nucleation of a magnetic domain during reversal. Here, we experimentally demonstrate ultrafast SOT-induced magnetization switching dynamics of a ferromagnetic sample with no incubation delay by avoiding the nucleation process and driving the magnetization coherently. We employ an ultrafast photo-conducting (Auston) switch and a co-planar strip line to generate and guide $\sim ps$ current pulses into the heavy metal/ferromagnetic layer stack and thereby induce ultrafast SOT. We then use magneto-optical probing to investigate the magnetization switching dynamics with sub-picosecond time resolution. Depending on the relative current pulse and in-plane magnetic field polarities, we observe either an ultrafast demagnetization and subsequent recovery along with a SOT-induced precessional oscillation (but no switching), or ultrafast SOT switching. In the case of switching, the zero-crossing of magnetization takes place in $\sim 70\ ps$, which is approximately an order of magnitude faster compared to previous studies. Complete switching occurs in $\sim 250\ ps$ and is limited by cooling back to room temperature via thermal diffusion to the substrate. We use a macro-magnetic simulation coupled with an ultrafast heating model to analyze the effect of ultrafast thermal anisotropy torque and current-induced torque in the observed dynamics. Good agreement between our experimental results and the macro-spin model shows that the switching dynamics are coherent and present no discernable incubation delay in this ultrafast switching regime. Our work suggests a potential pathway toward dramatically increasing the writing speed of SOT magnetic random-access memory devices.

# I.    Introduction

In recent years, spintronics[1–3] has gained considerable attention and shows great promise for low-power, non-volatile memory technology. While magnetic memory devices based on spin-transfer torque (STT) have already been introduced to the market[4–6], they still present some challenges such as device lifetime ($10^{10}$ write cycles limited by tunnel barrier breakdown), relatively slow switching speed and the write error rate (due to the stochastic thermal fluctuation induced switching initiation), which are affecting its performance compared to the state-of-the-art semiconductor memory devices[3,4]. Spin-orbit torque (SOT) based devices, on the other hand, are expected to largely overcome these problems. Current-induced SOT switching possesses significant advantages over current-induced STT switching as the former does not require the switching current to pass through the MgO tunnel barrier. In STT devices with out-of-plane magnetization (widely used due to favorable dimensional scaling properties), the torque is parallel to the magnetization of the free layer and requires thermal fluctuations to initiate the switching process leading to a stochastic incubation delay. On the other hand, SOT is orthogonal to the magnetization of the free layer, which is expected to provide "instant-on" switching torque. However, recent studies of SOT switching dynamics on the $\sim ns$ time scale in ferro/ferri-magnets also reveal an incubation delay, which is attributed to the stochastic nature of domain wall nucleation[7–9]. In most experiments, the operation speed of SOT devices, i.e. the observed switching timescale for the magnetic layer, has been limited by the rise time and width of the current pulse source (typically in the $\sim ns$ range)[9–13]. For comparison, a ferrimagnet[14,15] or hybrid ferrimagnet/ferromagnet structures[16–18] can be switched in a few $ps$ using the thermally activated helicity-independent all-optical switching mechanism. Such switching can also be manifested using $\sim ps$ current pulses obtained from appropriately designed, optically excited ultrafast photoconductive (Auston) switches[19]. Coherent reversal of in-plane magnetic dots in $\sim 200\ ps$ has been observed earlier utilizing a specially designed magnetic field pulse generator from two coupled Auston switches[20]. Recently, SOT-induced switching in a ferromagnet[21] has been demonstrated using $\sim 6\ ps$ current pulses from such an Auston switch embedded in a coplanar waveguide. In this work, a temperature-dependent Landau-Lifshitz-Gilbert (LLG) based macro-magnetic simulation[21] suggested a coherent rotation of magnetization with the switching timescale being an order of magnitude faster than conventional $\sim ns$ SOT switching events which are governed by domain wall nucleation and propagation. However, experimental measurements of the switching dynamics were not reported in that work.

Here, we generate $\sim 9\ ps$ current pulses from an Auston switch embedded in a coplanar strip line (CPS) waveguide, guide them into a magnetic heterostructure containing heavy metal and Co ferromagnetic layers, and study the resulting ultrafast SOT-induced switching dynamics. Observation of the final state of the magnetization after the injection of single $\sim ps$ pulses in the structure reveals that the switching depends on the relative direction of the current pulse and the external in-plane symmetry breaking magnetic field ($H_x$), fully consistent with the symmetries expected from SOT, as in previous results[12,21]. Time-resolved magneto-optical Kerr effect (MOKE) experiments allow us to observe the ultrafast magnetization dynamics as a function of the current pulse and in-plane field direction and amplitude. We observe ultrafast magnetization switching (zero-crossing) in $\sim 70\ ps$ after the current pulse excitation, even for our device size of $\sim 5\ \times\ 4\ \mu m^2$, which is an order of magnitude faster compared to previous SOT switching studies using nanometer device dimensions[7,9,12]. Full reversal is achieved in $\sim 250\ ps$, set by the time for the ferromagnetic layer to cool via heat diffusion to the substrate.

This switching speed cannot be understood in terms of current-driven domain-wall dynamics. We use a modified LLG equation macro-spin simulation including damping-like and field-like SOT (with corresponding spin-hall angles $\Theta_{DL}$ and $\Theta_{FL}$, respectively) and a one-temperature ultrafast heating model to explain the observed time-resolved dynamics. The simulation with a dominant $\Theta_{DL}$ qualitatively reproduces the experimentally observed magnetization dynamics and provides important insight into the mechanism of the switching phenomenon. The ultrafast switching is attributed to the coherent rotation of magnetization by a large damping-like torque ($\tau_{DL}$) and ultrafast thermal anisotropy torque ($\tau_{TAT}$) of a softened magnet due to ultrafast Joule heating by the $\sim ps$ current pulse.

## II.    Sample Structure

We used D.C. magnetron sputtering to grow a magnetic multilayer structure containing a $1\ nm$ Co film with perpendicular magnetic anisotropy on top of a low temperature grown GaAs substrate and patterned magnetic device regions with dimensions of $\sim 5\ \times\ 4\ \mu m^2$. The ferromagnet has a coercivity of $\sim$200 Oe (see section 1 of the Supplemental Information). We then fabricated a CPS microwave waveguide structure (as shown in Fig. 1a) with the magnetic device embedded in both conductor lines using ultraviolet lithography and electron-beam evaporation. The multilayer stack structure is: Ta(5)/Pt(4)/Co(1)/Cu(1)/Ta(4)/Pt(1), where the thickness of each layer is given in nanometers in the subscript and schematically shown in the Fig. 1b. The CPS consists of two Ti(20nm)/Au(300 nm) electrode lines which connect the Auston switch with the two magnetic samples, such that the current flow is along opposite directions through these two lines. The optical image of the magnetic microdots embedded in the CPS waveguide is shown in Fig. 1c. The light contrast is the CPS line, and the dark contrast shows the magnetic structure. The details of the sample structure and the waveguide design can be found elsewhere[19,21]. The amplitude and the temporal shape of the current pulse were measured using a THz microprobe from Protemics GmbH, and we observe a current pulse of ~9 $ps$ full-width half-maximum (FWHM) as shown in the top inset of figure 1. Details of the electrical characterization of the Auston switch are given in section 2 of the Supplemental Information and are similar to previous reports[19,21].

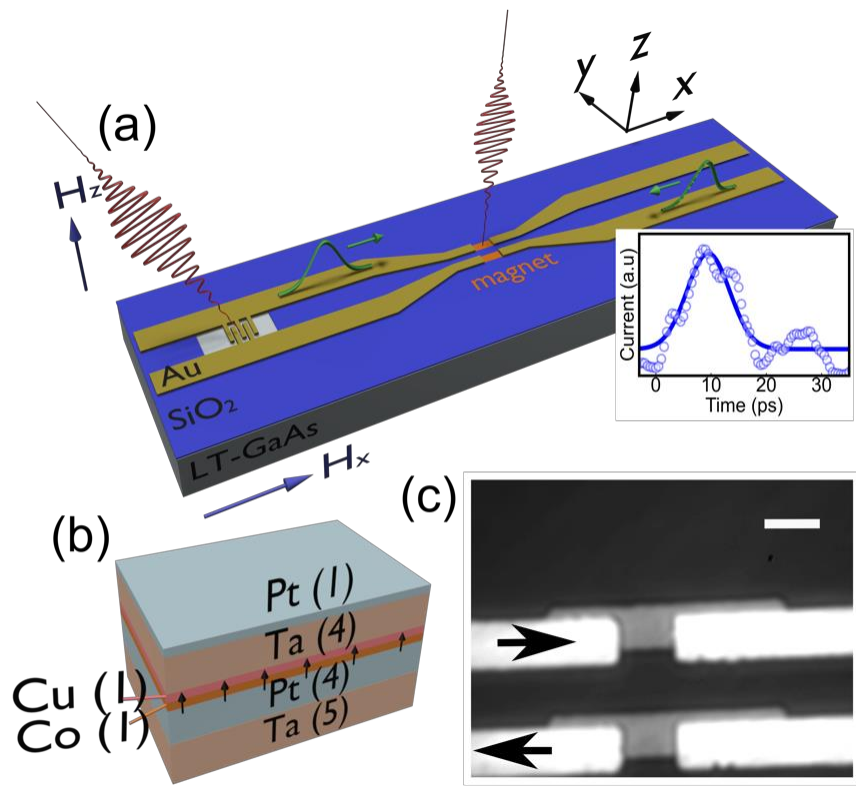

Fig. 1. (a) Schematic diagram of the Auston switch embedded in the CPS along with the pump and probe laser pulse and the ~$ps$ current pulse propagation, while the bottom right inset shows measured current pulse (blue dot) using the Protemics tip and a Gaussian fitting (blue line) gives us ~9 $ps$ full width at half-maximum. (b) The schematic of the cross-sectional view of the magnetic multilayer where the black arrows denote perpendicular magnetic anisotropy in the 1 $nm$ Co layer and (c) the optical image of the magnet embedded in the CPS waveguide structure. The black arrows designate the ~$ps$ current pulse propagation direction for a positive bias across the Auston switch. The white scale bar is 5 $\mu m$.

The MOKE images of the magnetic microdots after exciting them with single-shot electrical pulses in the presence of 1600 $Oe$ in-plane magnetic field is shown in section 3 of the Supplemental Information. These images indicate that the final magnetic states of the microdots are independent of the initial states and depend only on the relative orientation of the current pulse and in-plane magnetic field as expected from the symmetries of the SOT in the magnetic heterostructures[12,21].

## III.    Time-Resolved Magnetization Dynamics

We measured the ultrafast magnetization dynamics using a conventional time-resolved pump-probe polar MOKE setup[18,19]. We generated the $\sim ps$ current pulse by focusing a $\sim 100\ fs$-long laser pulse centered at $800\ nm$, with $0.28\ \mu J$ energy, from a $252\ kHz$ regeneratively amplified Ti:sapphire laser on the $\pm\ 50\ V$ biased Auston switch (with interdigitated electrodes and an overall area of $100\ \times\ 75\ \mu m^2$). A weaker, synchronized $800\ nm$ wavelength pulse with $0.13\ nJ$ of energy, called the probe, was focused through a $50X$, long working distance microscope objective (from Mitutoyo) to $\sim 1.6\ \mu m$ diameter on the magnetic sample. The magnetization dynamics were measured in a stroboscopic fashion, by varying the relative time delay between the pump and probe pulses using a motorized delay stage, and time-averaging the MOKE signal for each delay.

### A.  Ultrafast Magnetization Dynamics using Current Pulses without in-plane Magnetic Field

The current pulse-induced magnetization dynamics in the absence of $H_x$ are shown in Fig. 2a in black circles.  We observe a $\sim 30\ \%$ drop in the out-of-plane component of the magnetization due in part to ultrafast heating induced by the $\sim ps$ current pulses. The magnetization dynamics in the ferromagnetic microdot are further driven by the ultrafast SOT generated by the propagation of the $\sim ps$ current pulses through the heavy-metal layer in the heterostructure (in the absence of $H_x$). We use a $\sim 3.7\ mJ/cm^2$ incident optical fluence on the Auston switch with a bias voltage of $+50\ V$ to generate the current pulse with $\sim 9\ ps$ FWHM as shown in Fig. 1 and assume full saturation of the Auston switch at this optical fluence. Increasing the bias voltage beyond $50\ V$ caused damage to the Auston switch. We use a macro-spin model that includes ultrafast SOT and the effects of ultrafast temperature change, as presented in an earlier work[21], to fit the dynamics (for details see section 4 of the Supplemental Information). As the electrons and phonons can safely be estimated to be in thermal equilibrium during the entire process due to the $9 ps$ duration of the current pulse, we have simulated the ultrafast thermal change via a one-temperature model. The fit is shown by the dashed black line in Fig. 2a considering damping-like and field-like spin Hall angles of $\Theta_{DL} = 0.2$ and $\Theta_{FL} = 0.04$ respectively. A $9\ ps$ current pulse with a density of $7.6 \times 10^{12}\ A/m^2$ and $+300\ Oe$ out-of-plane bias magnetic field ($H_z$) is used in the simulation. The rest of the parameters are provided in section 4 of the Supplemental Information. The evolution of temperature as calculated from the one-temperature model and the corresponding change in the magnetization $\left( M_s(T) = M_s(0K)\left(1 - \frac{T_c}{T}\right)^{1.7} \right)$ as a function of time is plotted in Fig. 2b. The fitting suggests an ultrafast thermally assisted demagnetization close to $\sim 24\,\%$ and an associated maximum temperature rise of $\sim 460\,K$, as shown in Fig. 2b in blue and red lines, respectively. Hence the experimentally observed magnetization drop (of $\sim 30\,\%$) in Fig. 2a is dominated by the ultrafast heating in our sample. In the ultrafast SOT-induced switching measurements in the presence of $H_x$, as presented in the subsequent section, the contribution from the ultrafast thermal demagnetization remains the same. A separate measurement of the ultrafast demagnetization due to direct optical excitation of the magnetic sample by a $\sim 100\,fs$ laser pulse is shown in section 5 of the Supplemental Information, where we have observed $\sim 50\,\%$ magnetization change due to an absorbed laser fluence of $\sim 0.3\,mJ/cm^2$.

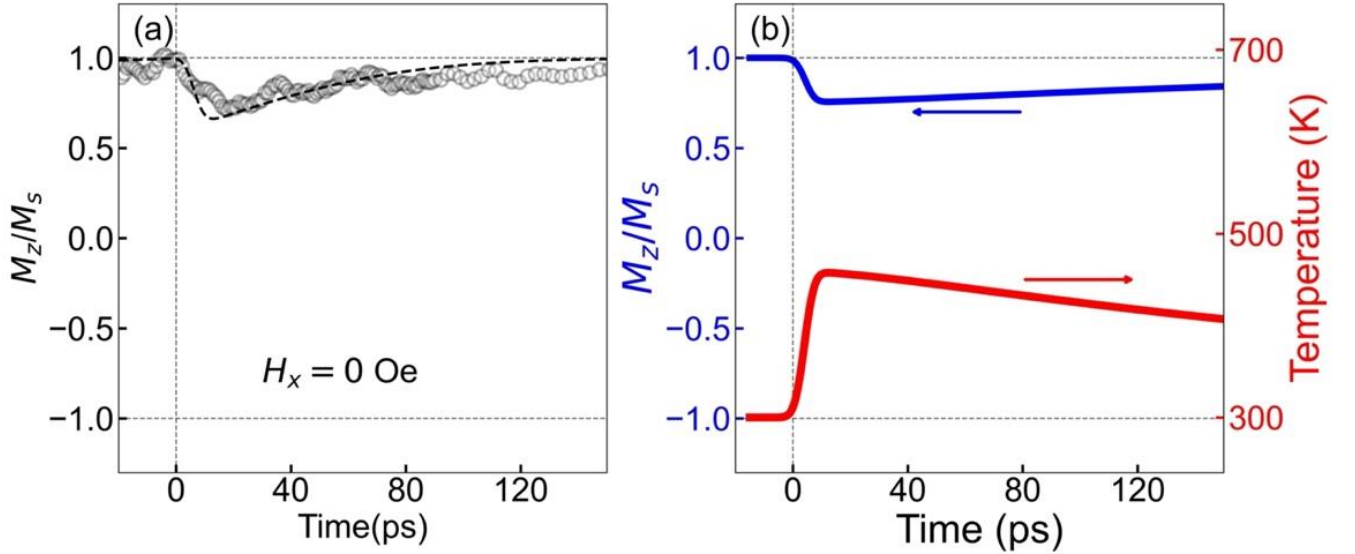

Fig. 2. (a) The time-resolved ultrafast demagnetization dynamics (in the absence of any in-plane magnetic field) using the $\sim ps$ current pulse excitation is shown in black circles. The dashed black line shows the fitting considering a macroscopic Landau-Lifshitz-Gilbert (LLG) model with $\theta_{DL} = 0.2$; $\theta_{FL} = 0.04$ and including ultrafast thermal effect due to the $9\,ps$ current pulse with a current density of $7.6 \times 10^{12}\,A/m^2$. (b) Simulated time-resolved demagnetization (blue) and the corresponding evolution of temperature increment (red) as calculated from the one-temperature model.

## B. Ultrafast Magnetization Switching

SOT-induced magnetization switching in our magnetic structures is achieved in the presence of $H_x$, as is observed from the single-shot MOKE micrographs shown in section 3 of the Supplemental Information. We observe the ultrafast SOT-induced magnetization switching dynamics via time-resolved MOKE starting from both a positive and negative magnetic saturation state, and for different in-plane field and current directions, as shown in Figs. 3a and b. We perform the experiments with an out-of-plane field $H_z$ (for details see section 1 in the Supplemental Information) above the coercivity of the sample to restore the ferromagnet to the same saturated initial state prior to each repetition of the $\sim ps$ excitation pulse. Starting from positive magnetic saturation in the presence of a positive $H_x$, we notice a modified initial drop in the out-of-plane magnetic component relative to the zero in-plane field case (which is discussed earlier in Fig. 2b) due to a negative $\sim ps$ current pulse. The corresponding remagnetization, as shown in Fig 3a in a solid golden line, is accompanied by precessional dynamics with two prominent magnetization oscillations (one smaller oscillation peak at $\sim 30\ ps$ and another larger oscillation peak at $\sim 60\ ps$). Similar dynamics are observed starting from a negative magnetic saturation in the presence of a negative $H_x$, as shown by the solid orange line in Fig. 3b. These oscillations are due to the damping-like torque ($\tau_{DL}$), which sharply tilts the magnetization in the opposite direction before the ultrafast thermal anisotropy torque ($\tau_{TAT}$) kicks into the oscillatory dynamics. The effect of the $\tau_{DL}$ vanishes after the finite duration of the $\sim ps$ current pulse, and the subsequent magnetization dynamics are then driven by $\tau_{TAT}$. The thermal heating by the $\sim ps$ current pulse creates a time-dependent change not only in the magnetization amplitude but also in the anisotropy field (for details see section 4 of the Supplemental Information). The tilted magnetization and the effective anisotropy field (time-dependent anisotropy field in combination with a constant external $H_x$ and $H_z$) become non-colinear and create a torque that lasts much longer than the current pulse duration. This is defined as $\tau_{TAT}$, whose amplitude depends both on $H_x$ and ultrafast thermal demagnetization. The oscillations get damped within $\sim 250\ ps$ due to the large Gilbert damping of the material[21], and it aligns the magnetization along the effective magnetic field at a longer timescale due to the Gilbert damping torque ($\tau_{GL}$). We note that the dynamics are not perfectly symmetric when starting from positive and negative saturations. This can be due to multiple reasons such as i) anisotropic current pulses in the CPS lines or/and ii) a difference in the effective in-plane magnetic field between positive and negative orientation (due to non-idealities in the external magnet structure used to apply the in-plane and out-of-plane fields) might lead to asymmetries in the anisotropic ultrafast torques. The effect of the positive $\sim ps$ current pulse, starting from positive magnetic saturation in the presence of a positive $H_x$, is shown in teal solid lines in Fig. 3a. Again, similar dynamics are observed starting from a negative magnetic saturation in the presence of a negative $H_x$, as shown by the solid blue line in Fig. 3b. We distinguish a sharp drop in the out-of-plane magnetization, where $\tau_{DL}$ aids $\tau_{TAT}$. As the out-of-plane magnetization crosses the equator (x-y plane), it is still driven by $\tau_{TAT}$ before the strong $\tau_{GL}$ settles the magnetization along the opposite saturation direction. We obtain a zero-crossing in $\sim 70\ ps$ and a nearly complete switching (magnetization reaches $\sim 75\ \%$ in the opposite direction[22]) in $\sim 250\ ps$. Using the experimentally obtained $9\ ps$ FWHM of the current pulse in our macro-spin simulation, we observe excellent agreement with the experimental data. The simulated dynamics are shown by the dashed lines in Fig. 3 using $\Theta_{DL} = 0.2$ and $\Theta_{FL} = 0.04$, for a $9\ ps$ current pulse with a current density of $\pm\ 7.6 \times 10^{12}\ A/m^2$ in the presence of a $\pm 1600\ Oe\ H_x$ and $300\ Oe\ H_z$. We notice that the magnetization doesn't recover to its full reversed saturation value after switching even after $600\ ps$, as shown in the long time-delay scans (Fig. S6 in Supplemental Information). This is attributed to the time required for the device to fully recover to room temperature via heat diffusion into the substrate in our sample.

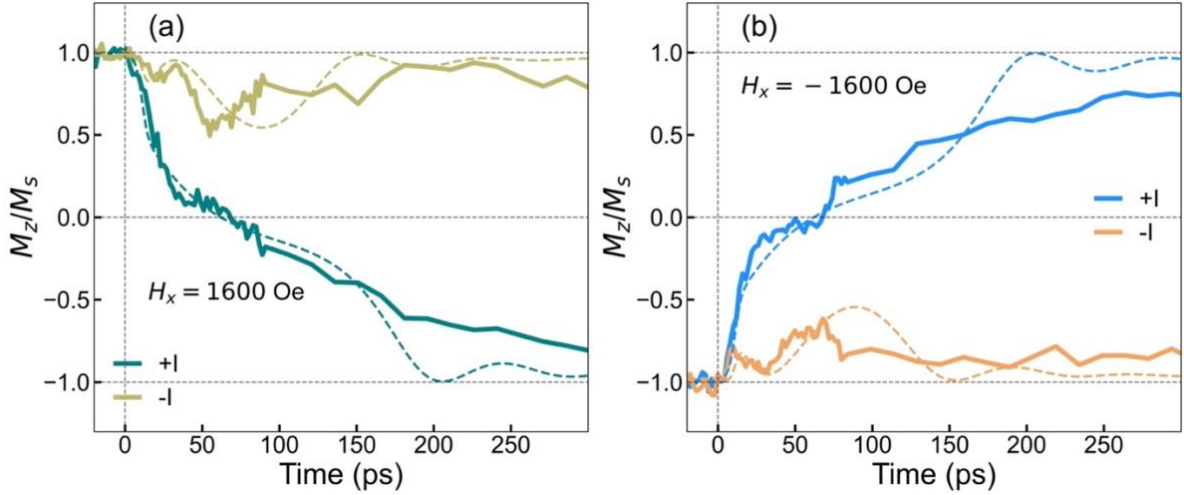

Fig. 3. The current pulse-induced time-resolved magnetization dynamics measured in the presence of (a) positive and (b) negative in-plane symmetry-breaking magnetic field and reversing the direction of the current pulses starting from positive and negative magnetic saturation. The dotted lines show the theoretical analysis by solving a macroscopic LLG equation with damping-like ($\Theta_{DL} = 0.2$) and field-like ($\Theta_{FL} = 0.04$) SOT terms and combining ultrafast one-temperature heating model due to a $9\ ps$ current pulse with a current density of $\pm\ 7.6 \times 10^{12}\ A/m^2$.

We show the simulated magnetization dynamics for a positive (negative) current pulse with increasing pump energy on the Auston switch in the presence of a constant $H_x$ in section 7 of the Supplemental Information. The zero-crossing time and the overall shape of the dynamics don't change with the increasing pump energy on the Auston switch. This confirms that the Auston switch is already operating in its saturation at $0.28\ \mu J$ pump energy.

The FWHM of the current pulse obtained from the Auston switch can vary between $\sim 6 - 10\ ps$ due to electrode geometry. Reliable ultrafast SOT-induced switching with $\sim 6\ ps$ current pulse has already been demonstrated earlier[21]. In the macroscopic simulation, we have varied the current pulse width between $6 - 15\ ps$ and observed consistent SOT-induced switching within $\sim 100\ ps$. The switching mechanism and the underlying physics remain the same within the range of the simulated current pulse width. However, the required current density reduces significantly with increasing FWHM as shown in Section 2 of the Supplemental Information. This would be important in a memory device application where the smaller peak current density can easily be obtained from a smaller transistor.

### C. Ultrafast Magnetization Switching Mechanism

Most of the previously reported SOT-induced switching experiments were performed with $\sim ns$ current pulses where the switching dynamics are governed by domain wall nucleation and propagation[7–9,11,13,23]. The typical SOT-induced domain wall velocities are in the range of $0.1 - 0.5\ km/s$ in the commonly-used ferromagnets like Co or CoFeB[9,12,13], which can increase to $\sim 5\ km/s$ in Gd-based ferrimagnets near compensation or in special ferromagnets with ultrasmall Gilbert damping[11,24] as observed in earlier studies. Considering a $\sim 250\ ps$ switching time, we calculate an unrealistic domain wall velocity (exceeding the Walker breakdown[24]) of $\sim 25\ km/s$ in our Co microdot with $\sim 5\ \times 4\ \mu m^2$ dimensions. Furthermore, classical domain wall nucleation is generally considered to be a stochastic process with an average nucleation time of the order of $1\ ns$ or greater[8,9,13,23]. We observe essentially no incubation time in our experiments. We do not resolve single-shot dynamics, as observed in some recent studies[8,9]. However, a random incubation delay of $\sim ns$ duration would manifest as a stretched-out rise time of the averaged transient signal, which we don't observe in our experiment. It has been previously envisioned that the incubation time to nucleate a domain wall can be significant or even completely suppressed by a strong perturbation with shorter current pulses, which is called an intrinsic SOT switching regime[8,12]. Simulations show that in this regime, SOT itself is sufficient to take the magnetization away from the equilibrium position to nucleate a domain wall without the need for any thermal energy (thermal rise time $\sim 2\ ns$)[8]. However, our measurement scenario is completely different as we use a $\sim ps$ current pulse and it induces a strong and ultrafast thermal change responsible for the demagnetization occurring in tens of $ps$ as observed in Fig. 2. The magnetization responds almost instantaneously to the applied $\sim ps$ current pulse with no incubation delay. Hence, we are operating in an intrinsic regime that is aided by ultrafast current-induced heating.

For a negative current pulse, we observe an ultrafast demagnetization along with the effect of $\tau_{DL}$, which tilts the magnetization in the opposite direction (compared to $\tau_{TAT}$), and $\tau_{TAT}$ initiates strong magnetization oscillations. As the anisotropy field ($H_{ani}$) is a function of temperature (which is again a function of time), the frequency of oscillation changes proportional to $2\pi/\gamma H_{ani}$, which is responsible for the two prominent oscillation peak centered at $\sim30\ ps$ and at $\sim60\ ps$ as shown in yellow (orange) solid lines in Fig 3a (Fig. 3b). At longer timescales, the $\tau_{GL}$ aligns the magnetization along its initial direction. When we apply a positive current pulse starting from positive saturation with the positive in-plane field, as shown in solid teal lines Fig. 3a, the current pulse partially demagnetizes the film on an ultrafast timescale[21,25] and at the same time $\tau_{DL}$ tilts the magnetization along the spin-polarization axis. In this case, the direction of $\tau_{DL}$ is opposite to that of the negative current pulse direction. Hence, the effect of $\tau_{DL}$ is enhanced by $\tau_{TAT}$. However, damping-like torque vanishes after the $9\ ps$ duration of the current pulse. Then the magnetization continues to be driven by $\tau_{TAT}$ and it crosses the equator plane. At a longer timescale, $\tau_{GL}$ relaxes the magnetization along the opposite direction[26]. Similar dynamics are observed for a positive current pulse starting from negative magnetic saturation, using a negative in-plane field, as shown by the solid blue line in Fig. 3b. Finally, the effect of field-like torque is observed to be minimal in our structures and is discussed in detail in the next section

This switching mechanism discussed here has been hypothesized in an earlier work[21] and these measurements now corroborate the hypothesis. We highlight that this precessional switching mechanism can only work if the heating, and change in effective field direction, happen at a faster timescale than the ferromagnetic resonance half-period. Otherwise, for slower heating, the magnetization would simply follow the effective field and could not swing across the equator plane unless the effective field itself did cross it. Our model mimics the zero-crossing time of $\sim70\ ps$ and shows close agreement with the shape of the experimental magnetization switching curve as shown in Fig. 3. In the simulated switching dynamics, the magnetization gets settled to its complete switched value faster than the experiment. In the simulation, we haven't considered the effect of the thicker oxide layer in between the magnetic layer and substrate, which might cause such a discrepancy.

The effect of $H_x$ on the ultrafast magnetization dynamics is shown in Section S8 of the Supplemental Information, where we observe that an $800\ Oe$ in-plane symmetry breaking field is insufficient to observe magnetization switching with the same current density. For the positive $\sim ps$ current pulse (Fig. S8b in Supplemental Information), $\tau_{TAT}$ and $\tau_{DL}$ work in tandem and try to drag the magnetization away from its equilibrium. However, with the smaller in-plane field, the magnitude of $\tau_{TAT}$ is smaller. Hence, to induce complete magnetization switching, a larger current density or in-plane magnetic field is required.

### D.  Simulated effect of Damping-like Torque and Current Density

The theoretical analysis for different values of $\Theta_{DL}$ (at a fixed $\Theta_{FL} = 0.04$) on the ultrafast magnetization dynamics is shown in Fig. 4a using the $9\ ps$ current pulse with a current density of $7.6 \times 10^{12}\ A/m^2$, and the corresponding three-dimensional magnetization trajectories are shown in Fig. 4b. The effect of the damping and the thermal anisotropy torques works in the same direction for positive and the opposite direction for negative current pulses. With a positive current pulse (shown in the solid lines in Fig. 4a-b), for the smallest damping-like SOT term ($\Theta_{DL} = 0.15$), the out-of-plane magnetization tilts sharply, however, together with $\tau_{TAT}$, it is still insufficient to produce enough torque for the magnetization to cross the equator and ultimately it settles along its initial direction. We observe magnetization switching for a larger $\Theta_{DL}$, and the zero-crossing times become faster with increasing $\Theta_{DL}$. For negative current pulses (as shown by the dotted lines in Fig. 4a-b), $\tau_{DL}$ tilts the magnetization in the opposite direction, which is observed by the initial change in the magnetization, and its amplitude increases with increasing $\tau_{DL}$. Now, the $\tau_{TAT}$ working in the opposite direction initiates the oscillations and we observe increased oscillation radius with increasing $\tau_{DL}$ from the dotted lines in Fig. 4b. The oscillations get damped by the strong $\tau_{GL}$ at longer timescales.

The effect of different current densities at a fixed value of $\Theta_{DL} = 0.2$ and $\Theta_{FL} = 0.04$, is shown in Fig. 4c along with the three-dimensional magnetization trajectories shown in Fig. 4d. Magnetization dynamics at relatively low current densities are accompanied with either slower or no-switching. With increasing current density, the effect of the thermal anisotropy torque becomes larger as the ultrafast heating gets stronger, and we observe faster zero-crossing for positive $\sim ps$ current pulses. The effect of the larger $\tau_{TAT}$ is also reflected in the increased oscillation amplitudes and reduced frequency $\left( \propto {2\pi}/{\gamma H_{ani}} \right)$ along the negative current direction before the magnetization settles towards its initial saturation. The value of $\Theta_{DL}$ and $\Theta_{FL}$ depends on the heavy-metal layer and the ferromagnet/heavy-metal interface. For our magnetic stack, the effect of $\tau_{FL}$ is minimal as observed in the previous studies[21,27] and also corroborated by our simulation (see section 9 of the Supplemental Information). After zero-crossing, the magnetization stabilization gets delayed with increasing field-like torque. The coherent nature of the observed ultrafast magnetization reversal is well-established with our macro-spin model, and it suggests the possibility of achieving a much faster ($\sim20\ ps$) zero-crossing time.

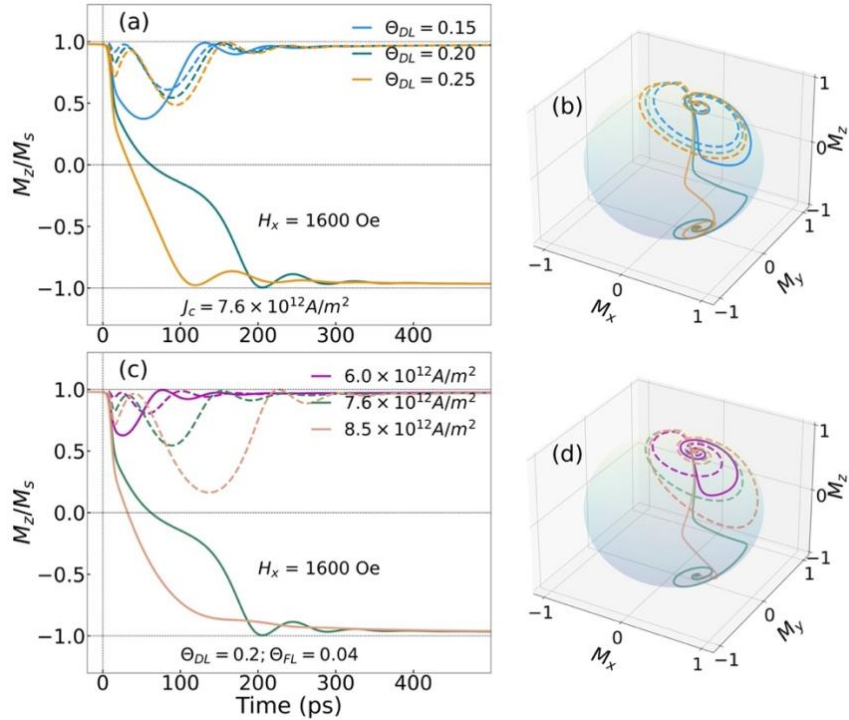

Fig. 4. The theoretical magnetization dynamics by solving LLG-SOT equation with the ultrafast one-temperature model (a) for different values of damping-like ($\Theta_{DL}$) torque (at a fixed $\Theta_{FL} = 0.04$) using a 9 $ps$ current pulse with a fixed current density of $7.6 \times 10^{12}\ A/m^2$, and (b) the corresponding three-dimensional dynamics of the magnetization vector for three different damping-like torque values. (c) The magnetization dynamics for different current densities at a fixed value of $\Theta_{DL} = 0.2$ and $\Theta_{FL} = 0.04$, in the presence of a 1600 $Oe$ symmetry-breaking in-plane field and the (d) corresponding three-dimensional dynamics of the magnetization vector. The solid (dotted) lines show the magnetization dynamics for positive (negative) current pulses.

We performed microscopic spin simulations with UBERMAG[28] (OOMMF based simulation technique in python framework) due to a current pulse with 9 $ps$ FWHM at 0 K and without introducing any ultrafast heating. This is described in detail in Section 10 of Supplemental Information. In the absence of $\tau_{TAT}$, we don't observe magnetization switching with a current density of $7.6 \times 10^{12}\ A/m^2$. The minimum current density required to obtain switching is $16.5 \times 10^{12}\ A/m^2$, when the magnetization drops close to zero within the full duration of the current pulse and it crosses the equator plane within $\sim40\ ps$. We don't observe any domain wall nucleation within this time frame and the switching occurs via a coherent rotation of the magnetization of the whole sample without any noticeable incubation delay. The microscopic simulation emphasizes the need for $\tau_{TAT}$ in obtaining ultrafast switching at a smaller current density and establishes the coherent switching mechanism even in the micro-spin picture with $\sim ps$ current pulses.

However, when we excited the magnet with a $1\ ns$ square wave electrical pulse, we do observe domain walls-driven magnetization switching at a smaller current density.

### E. Optimal Current and Energy Density

The FWHM of the current pulse sets an upper limit of the $RC$ constant of the Auston switch ($RC < 9\ ps$) and we can estimate a maximum capacitance ($C_{max}$) of $\sim 12.8 \times 10^{-14}$ F[21] (as the resistance is $\sim 70\ \Omega$). This translates to a maximum of $\sim 160\ pJ$ maximum electrical energy $\left(E_{max} = \frac{1}{2}C_{max}V^2\right)$ stored in the Auston switch. Assuming all this energy is being transferred to the $\sim 9\ ps$ Gaussian current pulse and ultimately delivered to the magnetic microdot ($\sim 5 \times 4\ \mu m^2$), a maximum electrical energy density of $\sim 0.81\ mJ/cm^2$ could be delivered to the sample. The maximum current density ($J_{c,max}$) that can be obtained from the electrical energy, is $\sim 9.6 \times 10^{12}\ A/m^2$ $\left(E_{max} = \int J_{c,max}^2(t)\rho dt\right)$. The actual current density in the device should be smaller than this value due to various losses, which is consistent with the current density obtained in the macro-spin simulations. Our simulated current density $\left(J_c = 7.6 \times 10^{12}\ A/m^2\right)$ from the macroscopic model is an order of magnitude larger when compared to $\sim ns$ current pulse-induced SOT switching measurements[8,29]. However, this current density is needed only for $\sim 9\ ps$ (instead of up to tens of $ns$), and hence the corresponding maximum energy density ($\sim 0.75 \times J_c^2(t)\rho\tau$) is $\sim 0.51\ aJ/nm^3$ is small. The required energy would reduce only to $\sim 4\ fJ$ in an actual device with a $20\ nm^3$ size magnetic dot[21], which is significantly smaller than the reported switching energy values when using $\sim ns$ current pulses[11]. The final SOT-MRAM cell size is limited by the driver transistor size as a larger current can only be delivered by a larger transistor, which will ultimately increase the array size or reduce the memory density. Hence, significant improvement is required to obtain faster switching at an optimal current density and we believe heavy-metal layers or topological insulators with much larger SOT efficiency can be an important step forward to reduce the required current density. SOT efficiency can also be larger than 1[30,31] in specially designed current-carrying materials, which promises an even smaller threshold current density in an optimized magnetic stack. Our micromagnetic simulations estimated a much smaller current density when using a $sub-100\ ps$ current pulse, while maintaining the coherence of the magnetization switching and an ultrafast magnetization zero-crossing. Some recent studies reveal that the use of a small STT current on top of the SOT current can significantly reduce the switching current density[32]. The ultrashort current pulse-width may also help in enhancing the device lifetime, which can be damaged by electromigration and self-heating induced diffusion, as has been observed in recent theoretical works comparing $100 \ ns$ and $1 \ ns$ current pulses[33].

## IV.  Conclusion

We have demonstrated time-resolved ultrafast SOT-induced magnetization dynamics in a $1 \ nm$ Co film using $\sim ps$ current pulses generated in a biased Auston switch patterned into a CPS. Single-shot MOKE imaging experiments confirm that the magnetization switching depends on the relative orientation of the symmetry-breaking in-plane magnetic field and the current pulse direction, consistent with the expected symmetries of SOT switching in our magnetic stack. The current pulse induces $\sim 24\%$ ultrafast demagnetization in the ferromagnetic microdot due to ultrafast thermal heating. The time-resolved magnetization dynamics reveal a zero-crossing of $\sim 70 \ ps$ starting from positive (negative) saturation in the presence of a positive (negative) in-plane magnetic field, respectively, due to a positive current pulse. The full switching takes about $\sim 250 \ ps$ with the full recovery of the switched magnetization amplitude limited by heat diffusion from the ferromagnetic microdot to the substrate. On the other hand, a negative current pulse doesn't induce switching for these combinations of initial saturation direction and in-plane field but generates ultrafast demagnetization together with SOT-induced oscillations, whose amplitude and period are governed by the magnitude of the damping and the temporal evolution of the effective anisotropy field. A macro-spin LLG simulation combined with ultrafast thermal response agrees well with the experimentally observed magnetization dynamics. The measured SOT-induced oscillations and the zero-crossing time have been well-reproduced with a dominant damping-like and a small field-like torque in the presence of an ultrafast thermal anisotropy torque in the macroscopic simulation. While previous SOT experiments with $\sim ns$ long current pulses confirm domain wall nucleation and propagation-driven magnetization dynamics, our present measurement with ultrafast SOT shows a magnetization reversal by a rapid coherent rotation of the partially demagnetized magnetic moments towards the switched direction. We hypothesize that this difference may be attributed to a combination of strong peak driving torque combined with the transient ultrafast heating and thermal anisotropy torque, which obviates the need for domain wall nucleation. This work points the way towards achieving integrated, on-chip SOT switching on sub-$100 \ ps$ timescales. Our work sets an important milestone in demonstrating the possibility of coherent ultrafast SOT switching and opens up a new realm of ultrafast magnetism, combining non-equilibrium heating with SOT effects. Future studies may lead to further enhancement in efficiency using optimized material stacks and drive current pulse widths.

## ACKNOWLEDGEMENTS

This work was supported by ASCENT, one of six centers in JUMP, a Semiconductor Research Corporation (SRC) program also sponsored by DARPA (instrumentation and data acquisition) and by the ANR Project ANR 20-CE24-0003 SPOTZ and the FEDER-FSE Lorraine et Massif Vosges 2014–2020. This work was also supported by the Director, Office of Science, Office of Basic Energy Sciences, Materials Sciences and Engineering Division, of the U.S. Department of Energy under Contract No. DE-AC02-05-CH11231 within the Nonequilibrium Magnetic Materials Program (MSMAG). We acknowledge support from the National Science Foundation Center for Energy Efficient Electronics Science and the Berkeley Emerging Technology Research (BETR) Center (instrumentation and data acquisition). We thank the French RENATECH network for its support. We acknowledge our fruitful discussion with Prof. Pietro Gambardella.

## Author Contributions

D.P. and A.P. designed the experiments with input from J.G., and J.B. A.L. grew the LT-GaAs substrates. M.H. optimized and grew the samples by sputtering. K.J. and E.D. fabricated the devices. D.P. and A.P. performed ultrafast SOT experiments and MOKE microscopy. A.R. and A.P. developed the macroscopic simulation method. D.P. performed the simulations after discussing with A.R., A.P., and J.B. D.P. developed and performed the microscopic simulation. D.P. characterized the Auston switches with the help of H.S. D.P. analyzed the experimental data with help from A.P., H.S., J.G., and J.B. D.P. wrote the manuscript with input from all authors.

## Competing Interests

The authors declare no competing interests.

# Supplementary Information

## Section 1: Hysteresis loops in the presence of an in-plane magnetic field

Due to the proximity of the in-plane ($H_{+x}$) and out-of-plane ($H_z$) external magnets in our experimental setup (as shown in Fig. S1a), the in-plane magnet provides a significant out-of-plane magnetic field as the field lines bend towards the core of the out-of-plane magnet.

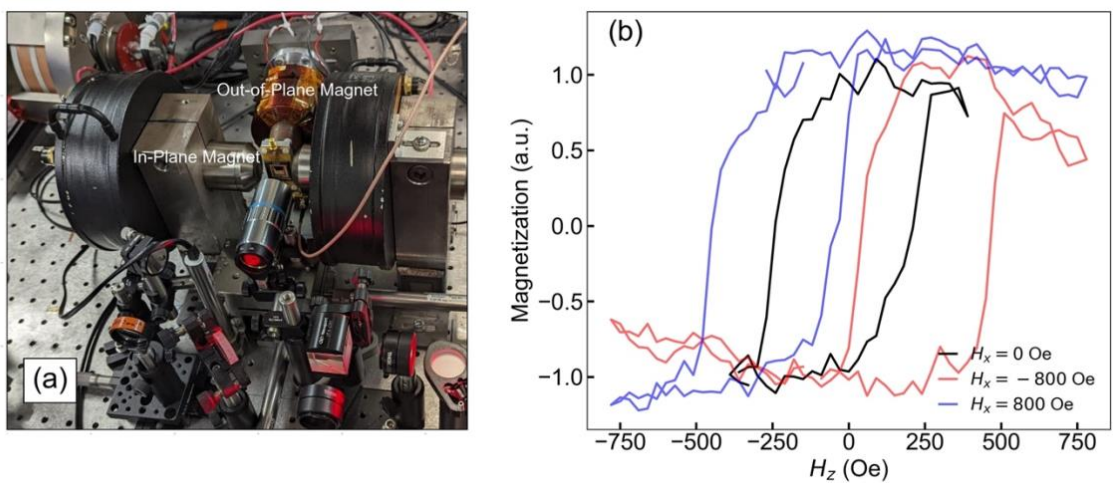

Fig. S1. (a) The proximity of in-plane and out-of-plane magnets and (b) the out-of-plane hysteresis loops in the presence of different in-plane magnetic field.

This effect can be observed from the shifts in the hysteresis loops of the Co microdot in presence of 0 and ± 800 Oe in-plane field as shown in Fig. S1b, where the out-of-plane magnetic field is plotted in the x-axis. As we observe from this figure, even an 800 Oe in-plane magnetic field has almost enough out-of-plane components to switch the magnet. Hence, during time-resolved measurement with 1600 Oe in-plane field, we haven't applied any out-of-plane magnet separately and used the out-of-plane component of the in-plane magnet to provide an out-of-plane saturation magnetic field. However, for the time-resolved measurements with an 800 Oe in-plane magnetic field, we have explicitly used the out-of-plane magnet to provide the out-of-plane saturation field. In all cases, a saturating out-of-plane field is required in the repetitive pump-probe time-resolved measurements to reset the magnet back to the same initial condition for each successive cycle.

## Section 2: Characteristics of the Auston switch and current-pulse width dependence of the ultrafast

**switching**

The $IV$ characteristics of the photoconductive (Auston) switch is shown in Fig. S2_1a. We use a $60\ mW$ pump power (at $252\ KHz$ repetition rate) to study the photo-current and the measured photocurrent is $\sim 10$ times larger than the dark-current at the largest bias voltage (i.e. $50\ V$). We highlight that the Auston switch gets saturated with $60\ mW$ pump power (at $252\ KHz$ repetition rate; $0.28\ \mu J$) and at a fixed voltage, the photocurrent doesn't increase much beyond this pump power. We use a THz probe tip from Protemics GmbH on top of the co-planar strip-line (CPS), which measures the THz field generated from the THz current pulse propagating through the CPS. The measured FWHM of the current pulse width is $\sim 9\ ps$.

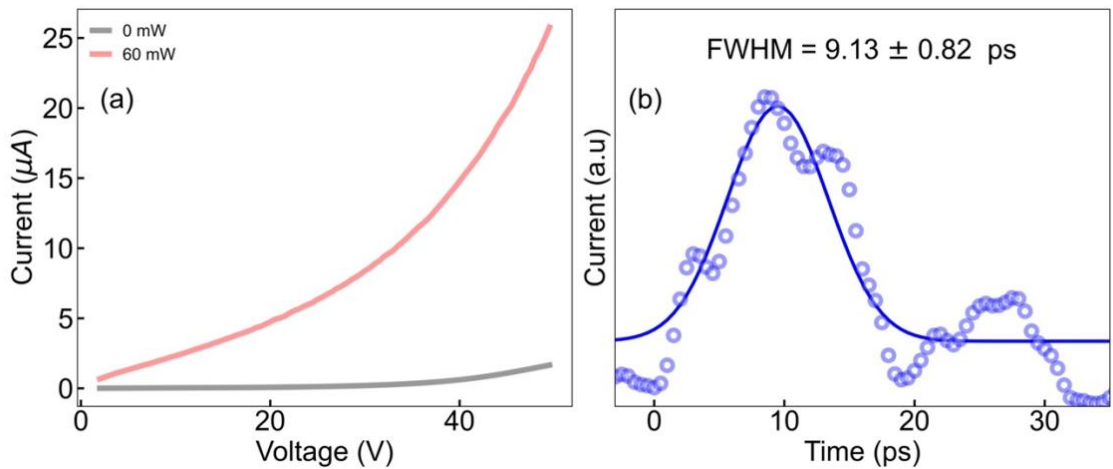

Fig. S2_1. (a) The dark-current (black) and photo-current (red) as a function of the applied bias voltage and (b) the measured current pulse from the Auston switch using a THz probe tip from Protemics GmbH is shown in blue dots. The solid blue line is a Gaussian fitting showing $\sim 9\ ps$ FWHM.

The FWHM of the current pulse depends on the Auston switch configuration, waveguide structure and the LT-GaAs substrate. In different types of Auston switches, we generally observe a current pulse width variation between $\sim 6 - 10\ ps$ using the amplified laser excitation. As expected, for the smaller current pulse width, one would require a larger current density to observe switching and vice-versa[4]. Using our microscopic simulation (details of which are given in section 4), we have calculated that with a $6\ ps$ and $15\ ps$ current pulse-width, we need $10 \times 10^{12}\ A/m^2$ ($6 \times 10^{12}\ A/m^2$) current density to observe a similar ultrafast SOT-induced magnetization switching which is shown as shown in Fig. S2_2. We always use an in-plane symmetry-breaking field of $1600\ Oe$ in these two simulations. The main takeaway is that the physics of the switching remains the same within a small variation in the current pulse width. The switching mechanism is still dominated by coherent rotation of magnetization without any noticeable incubation delay.

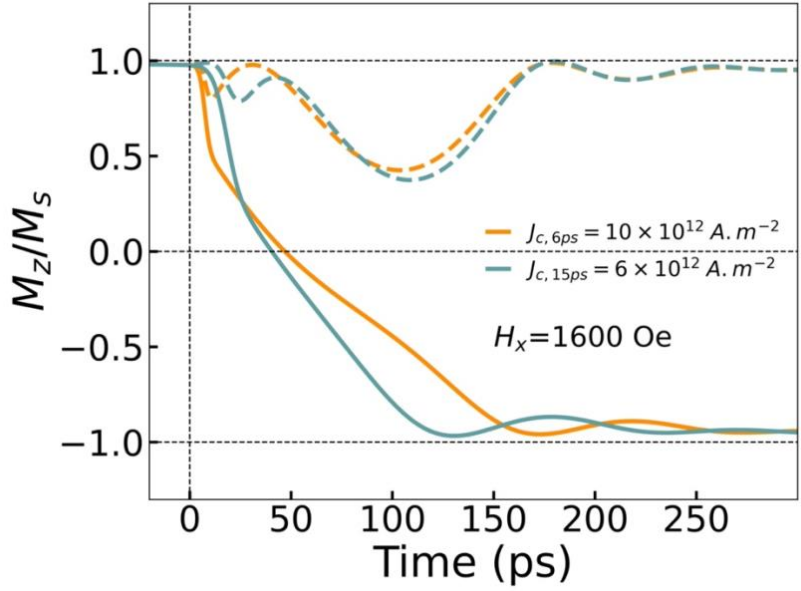

Fig. S2_2. The simulated magnetization dynamics using a 6 $ps$ and 15 $ps$ current pulse that require a current density of $10 \times 10^{12} \, A/m^2$ and $6 \times 10^{12} \, A/m^2$ respectively to induce ultrafast SOT induced magnetization switching. The solid (dotted) lines are obtained using a positive (negative) current pulse in the presence of 1600 $Oe$ in-plane symmetry-breaking magnetic field.

## Section 3: Single-shot MOKE switching images

In the presence of 1600 $Oe$ in-plane magnetic field, we acquire the first magneto-optical Kerr effect (MOKE) image after saturating the sample along $H_{+z}$ or $H_{-z}$ (i.e. its magnetization starts from $M_{+z}$ or $M_{-z}$ orientation, respectively) and then capture the second MOKE image after passing a single $\sim 9 \, ps$ current pulse (SS) through the electrodes (by single-shot optical excitation of the Auston switch under a 50 $V$ bias). The sign of the $\sim ps$ current pulses along the two electrodes is shown by the green arrows at the top and bottom rows of the schematic for positive and negative biases, respectively, across the Auston switch. The difference between these two images is shown in the first and second columns of Fig. S3 respectively. The black (or white) contrast signifies a complete switching of the magnet from $M_{+z}$ to $M_{-z}$ (or $M_{-z}$ to $M_{+z}$) magnetic saturation state and the grey contrast signifies no change in the magnetic state. In the top-left quadrant of Fig. S3, the top magnetic micro-dot switches, which means it prefers $M_{-z}$ orientation (shown by the black contrast) for the parallel combination of in-plane magnetic field and $\sim ps$ current pulse directions. However, the bottom microdot doesn't switch (shown by the grey color).

Similarly, if we start the experiment from $M_{-z}$ orientation, the top magnetic micro-dot doesn't switch, and the bottom microdot switches to $M_{+z}$ as shown by the white contrast. A similar switching phenomenon is observed by reversing the bias voltage of the Auston switch (bottom two micrographs of Fig S3). In this case, the final magnetization directions are opposite to the final states of the top two micrographs since the current directions have been reversed. These MOKE contrasts indicate that the final magnetic states of the microdot are independent of the initial states and depend only on the relative orientation of the current pulse and in-plane magnetic field as expected from the symmetries of the SOT in the magnetic heterostructures[12,20]. We repeated each measurement five times and observed the same final magnetic switched state. Lowering the bias voltage below $50\,V$, without increasing the in-plane magnetic field, results in no switching.

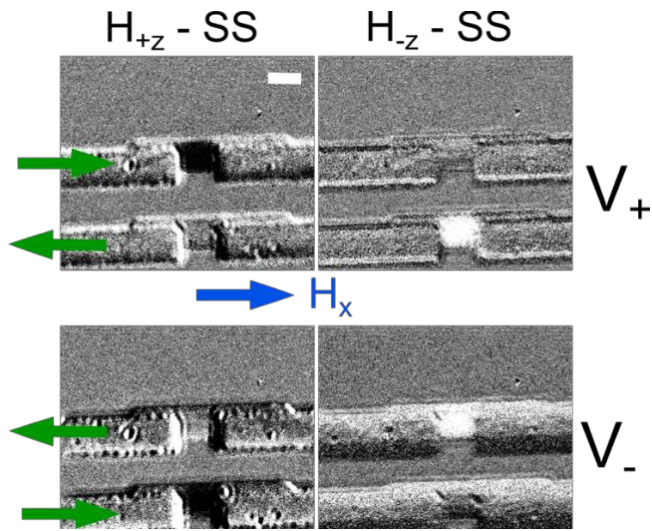

Fig. S3. The differential magneto-optical Kerr effect (MOKE) images showing the reversal of the magnetization due to a single current pulse (green arrow) in the presence of $1600\,Oe$ in-plane symmetry breaking magnetic field ($H_x$), under positive and negative bias voltages ($V_+$ and $V_-$) across the Auston switch, starting from a positive and negative magnetic saturation. We capture two MOKE images; 1) after saturating the magnets either along $H_{+z}$ or $H_{-z}$ direction and ii) after passing a single ~9 $ps$ current pulse (SS) and the subtraction of these two images produces the differential MOKE contrast, where if a switching event occurs from $H_{+z}$ to $H_{-z}$ (or $H_{-z}$ to $H_{+z}$), it appears as black (or white) contrast, and a gray represents no change in magnetization. The scalebar is $5\,\mu m$.

## Section 4: LLG Simulation with ultrafast heating

We have adopted the same technique used in one of our previous works to study SOT-induced magnetization dynamics[2]. The LLG differential equation for a ferromagnet in an external field can be extended by linearity to include the effect of the spin current as given below.

$$\frac{dM(t)}{dt} = -\gamma\mu_0\left(\vec{M} \times \overrightarrow{H_{eff}}\right) + \frac{\alpha}{M_s(T)}\left(\vec{M} \times \frac{dM(t)}{dt}\right) + \Theta_{FL}C_S\left(\vec{M} \times \vec{\sigma}\right) - \Theta_{DL}\frac{C_S}{M_s(T)}\left(\vec{M} \times \vec{M} \times \vec{\sigma}\right).$$

Here, γ, μ₀, H_eff, α, and σ are the gyromagnetic ratio, vacuum permeability, effective magnetic field, Gilbert damping, and spin-polarization direction, respectively. The torques on the magnetization due to the spin current's magnetic moment is modulated by the field-like spin Hall angle and the damping spin Hall angle. In the first two terms detailing the interaction of the ferromagnet with the external field, $\mu_0$ is the permeability of free space, $\gamma$ is the gyromagnetic ratio of electrons, $\vec{M}$ is the magnetization of Cobalt, $H_{eff}$ is the effective external field which consists of the external field, demagnetization field, and magneto-crystalline anisotropy, $\alpha$ is Gilbert damping parameter, and $M_s$ is the saturation magnetization. In the last two terms that incorporate the action of the spin currents, $\Theta_{FL}$ and is the field-like and anti-damping spin Hall angles, respectively, and $\sigma$ is the unit vector that points in the same direction as the spin current's magnetic moment while the magnitude of the torque's effect is absorbed into the $C_S$ variable, where, $C_S = \frac{\mu_B J_c}{q_e d_0 M_s}$, and $\mu_B$, $J_c$, $q_e$ and $d_0$ are the Bohr magnetron, time-varying current density, the elementary charge of an electron and the thickness of Cobalt respectively.

Now, as we already know, with $\sim ps$ current injection in a metallic magnetic stack, the thin films undergo non-equilibrium Joule heating, and with the peak electron temperature rising much higher than room temperature and thereby it shows ultrafast demagnetization[2,3]. Hence, we have considered the thermal evolution of the saturation magnetization and the anisotropy in the LLG simulation. If the metallic stack is assumed to be symmetric in two dimensions and varies only in one direction, the temperature response of the metal film can be estimated by solving the following one spatial dimension heat-diffusion equation.

$$C\frac{\partial T(x,t)}{\partial t} = \Lambda\frac{\partial^2 T(x,t)}{\partial x^2} + q(t).$$

Here, $T(x,t)$ is the temperature of the film, $C$ is the volumetric heat capacity, $\Lambda$ is the metal's thermal conductivity, and $q(t)$ is the external power per unit volume generated due to the electric pulse. Assuming Ohmic behavior, volumetric heating due to the current can be estimated using $q(t) = \rho_e J^2(t)$, where $\rho_e = 10\ \Omega nm$ is the measured resistivity across the stack and $J(t)$, is the current density. The volumetric heat capacity of the full metallic stack is assumed to be, $C = 2.0\ Jm^{-3}K^{-1}$, and the thermal conductivity of the stack is calculated using the Wiedemann-Franz law $\Lambda = \frac{L_0 T}{\rho_e} = 9\ Wm^{-1}K^{-1}$ where $L_0$ is the Lorenz number. We have used here a one-temperature model as we know the electron-lattice thermalization timescales in these materials are in the range of several $\sim ps$, when the system can effectively be described by a single temperature without introducing significant error. The evolution of saturation magnetization and anisotropy is given by the following equation:

$$M_s(T) = M_s(0K)\left[1 - \frac{T}{T_c}\right]^{1.7}$$

$$K_z(T) = K_z(0K)\left[M_s(T)\Big/M_s(0K)\right]^{3.0}.$$

Here, $T_c$ is the Curie temperature of Cobalt, $M_s(0K)$ and $K_z(0K)$ is the saturation magnetization at anisotropy constant at absolute zero. The value of $M_s(T = 300\ K)$ is fixed to $10^6\ Am^{-1}$ from magnetometry measurements.

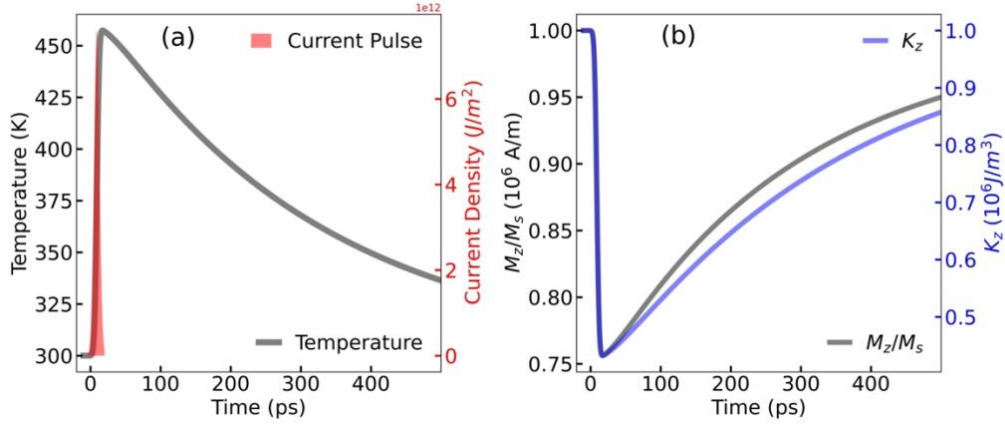

Fig S4. (a) Time-resolved temperature dependence due to the application of the ps current pulse depicted by the red filled region and (b) the time-resolved variation of the Magnetization (black line) and Uniaxial anisotropy (blue line).

The parameters used in this simulation are given in the following table.

| Parameter | Value |
|---|---|
| Volumetric Heat Capacity (C) | 2.0 x $10^6$ Jm$^{-3}$K$^{-1}$ |
| Interfacial Thermal Conductance (G) | 100 x $10^6$ Wm$^{-2}$K$^{-1}$ |
| Thickness of the Magnetic Layer ($d_0$) | 1 nm |
| Total Thickness (d) | 16 nm |
| Thermal Conductivity of the Metal ($\Lambda$) | 9 Wm$^{-1}$K$^{-1}$ |
| Electrical Resistivity ($\rho_e$) | 1.0 x $10^{-8}$ $\Omega$m |
| Current Density ($J_C$) | Variable (Am$^{-2}$) |
| FWHM of the Current Pulse | 9 ps |
| Saturation Magnetization ($M_S$) | $10^6$ Am$^{-1}$ |
| Anisotropy Constant ($K_z$) | $10^6$ Jm$^{-3}$ |
| Gilbert Damping ($\alpha$) | 0.22 |
| Currie Temperature ($T_c$) | 800 K |
| Out-of-plane Bias Magnetic Field ($H_z$) | 300 Oe |
| Symmetry-Breaking In-plane Magnetic Field ($H_{in}$) | 1600 Oe |

Table. S4. Parameters used in the macroscopic LLG simulation with ultrafast heating.

## Section 5: Ultrafast optical demagnetization

We use an incident optical pump fluence of $\sim 3.7\ mJ/cm^2$ to study the ultrafast optical demagnetization on a thin film of the sample as shown in the blue circles in Fig. S4. From the transfer function calculation, we found $\sim 8\ \%$ of the incident fluence ($\sim 0.3\ mJ/cm^2$) is being absorbed by the 1 nm thick Co layer. We experimentally observed an ultrafast thermal demagnetization of $\sim 50\ \%$. The maximum electron temperature rise is much larger due to the $\sim fs$ nature of the excitation (compared to a $\sim ps$ excitation) causing stronger demagnetization. We have used the following formulae[1] to fit the demagnetization curve:

$$\frac{\Delta M_Z(t)}{M_S} = \left[ \frac{A_1 \tau_e - A_2 \tau_m}{\tau_e - \tau_m} e^{-t/\tau_m} - \frac{\tau_m (A_1 - A_2)}{\tau_e - \tau_m} e^{-t/\tau_e} \right] \otimes [G(t)],$$

where $\tau_m$ and $\tau_e$ are the ultrafast demagnetization and fast remagnetization time, $A_1$ and $A_2$ are related to ultrafast demagnetization amplitude, and $G(t)$ is the Gaussian pump pulse with $100\ fs$ FWHM for optical and electrical excitation $\otimes$ denote a convolution between the two elements. We extract a $115 \pm 38\ fs$ demagnetization time and a remagnetization time of $1.34 \pm 0.16\ ps$ for optical excitation with the corresponding fitting shown as the dotted blue line in Fig. S4 using the above-mentioned model.

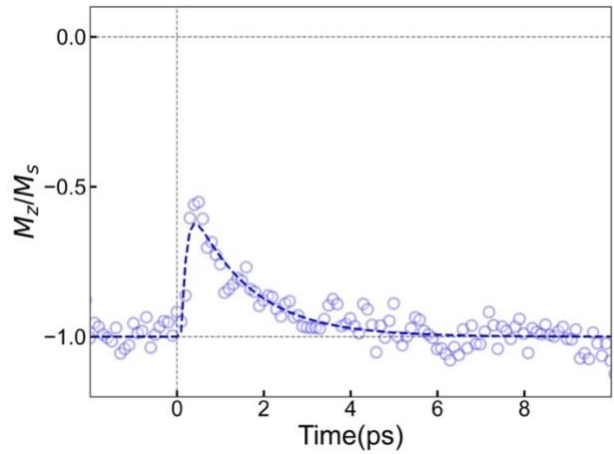

Fig S5. Ultrafast Optical Demagnetization due to a $\sim 100\ fs$ optical pulse as shown by the blue circles and the corresponding fitting with an empirical formula discussed in the text is given by the solid blue line

We note that the remagnetization time is much faster and doesn't go in line with the electrical magnetization measurement. We don't completely understand this but might be due to the different heat diffusion and slightly different magnetic properties between the ferromagnetic thin film (where optical demagnetization is measured) are measured and the ferromagnetic microdot (where electrical magnetization dynamics were studied).

## Section S6: Long time-delay measurements

The long time-delay scans of the time-resolved magnetization dynamics show that the magnetization doesn't recover completely after the current pulse excitation (measured till $\sim 600\ ps$). This suggests that the current pulse-induced thermal heating remains in the systems over this time. However, using the one-dimensional heating effect introduced in our macroscopic simulation model, the effect of the heating is not correctly reproduced, maybe the effect of individual interfaces along the multilayers structure and its diffusion through the MgO and LT-GaAs layer is required to match the experimentally observed long-time delay magnetization profile.

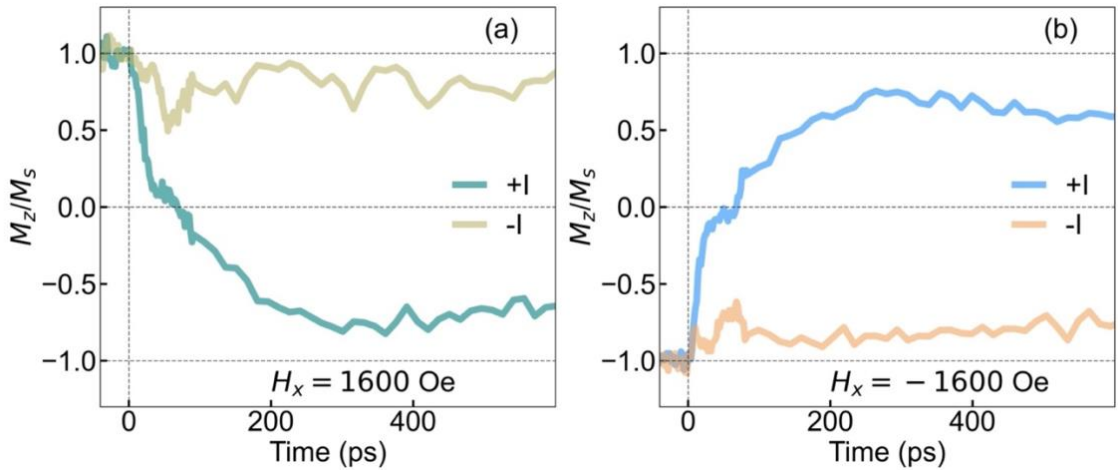

Fig. S6. The $\sim 9\ ps$ electrical pulse-induced time-resolved magnetization dynamics were measured in the presence of (a) positive and (b) negative in-plane symmetry breaking magnetic field and reversing the direction of the current pulses starting from positive and negative magnetic saturation. The solid (faint) lines denote the magnetization dynamics using positive (negative) $\sim 9\ ps$ current pulse.

## Section S7: The effect fluence on the Auston switch

We have used different fluence $(0.28, 0.30\ and\ 0.32\ \mu J)$ on the Auston switch at a fixed bias voltage of $\pm 50\ V$ and studied the magnetization dynamics as shown in Fig. S7 in the presence of a $1600\ Oe$ in-plane symmetry breaking magnetic field as shown in Fig. S7. The zero-crossing time and the overall shape of the magnetization dynamics don't change with increasing fluence on the Auston switch by $\sim 20\ \%$. This further strengthens our claim that we are operating the Auston switch at its saturation with $0.28\ \mu J$ pump energy on the Auston switch. Hence, the calculation of the capacitance of the Auston switch from the FWHM of the current pulse and the corresponding maximum energy stored in the Auston switch is valid throughout the magnetization switching experiments.

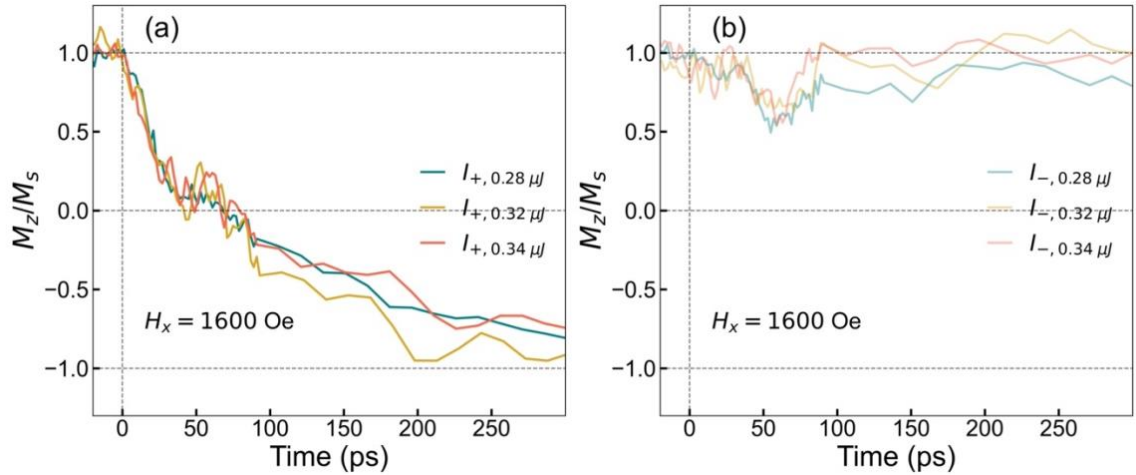

Fig. S7. The magnetization dynamics in the presence of 1600 $Oe$ in-plane magnetic field for (a) positive and (b) negative ~9 $ps$ current starting from a positive magnetic saturation for different pump energy on the Auston switch.

## Section S8: The effect of in-plane magnetic field

It is well-known that a symmetry-breaking in-plane magnetic field is required to for SOT-induced switching in perpendicular magnetic anisotropy magnets[2,5]. In our case, we need ~1600 $Oe$ in-plane magnetic field to observe switching as shown in the single-shot MOKE micrographs in Fig. S3 in the earlier section. When we used a smaller field, magnetization switching isn't observed as shown in Fig. S8a. With a lower in-plane field, a larger current density is likely needed to initiate switching.

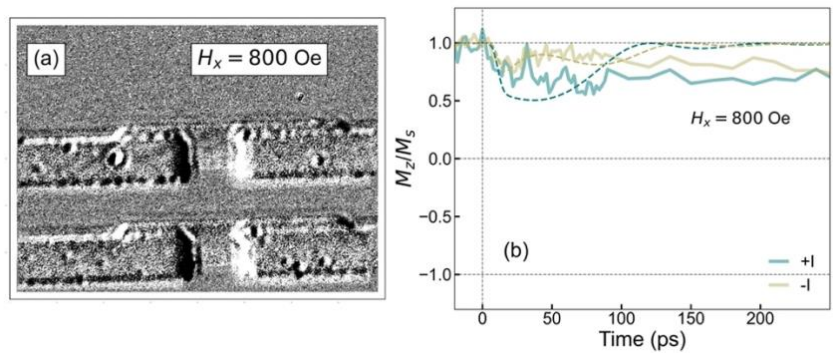

Fig. S8. (a) The single-shot MOKE micrograph starting from positive magnetic saturation in the presence of 800 $Oe$ in-plane symmetry-breaking field. The contrast in the magnetic sample doesn't change upon single-shot current pulse excitation which demonstrates no magnetization switching. (b) The corresponding magnetization dynamics are demonstrated. The simulated spectra shown in dotted lines for positive (green) and negative (yellow) current pulses show reasonable agreement with the experiment.

The time-resolved magnetization spectra are shown in Fig. S8b for both positive and negative current pulse direction while the dashed lines show the simulated results. The macroscopic simulation shows similar partial demagnetization behavior for a $\sim 9\ ps$ current pulse with a current density of $7.6 \times 10^{12} A/m^2$.

## Section 9: Effect of the field-like torque

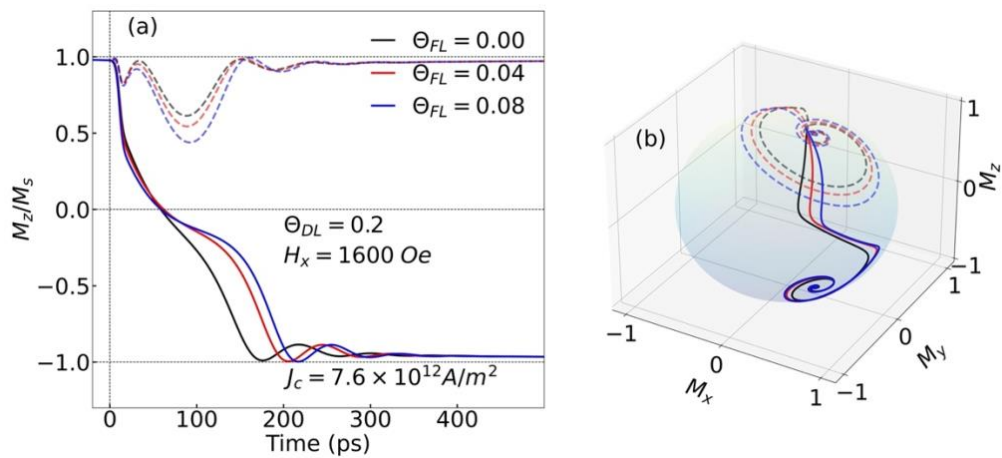

Fig. S9. Time-resolved magnetization dynamics in the presence of $1600\ Oe$ in-plane magnetic field due to a $\sim 9\ ps$ current pulse with a current density of $7.6 \times 10^{12}\ A/m^2$ for positive (solid) and negative (dotted) direction at three different field-like torque angles and the corresponding (b) three-dimensional magnetization profile.

We have shown the effect of the field-like torque angles while keeping the damping-like torque ($\Theta_{DL} = 0.2$) constant at a fixed current density ($J_c = 7.6 \times 10^{12} A/m^2$) of the $9\ ps$ current pulse in the presence of a positive $1600\ Oe$ in-plane symmetry breaking magnetic field for both positive (solid) and negative (dotted) current directions. The field-like torque increases the amplitude of the oscillation during remagnetization as shown in the dotted lines in Fig. S7. During magnetization switching, the magnetization crosses zero $\sim 70\ ps$ and it doesn't change with increasing $\Theta_{FL}$. However, magnetization settles to the negative saturation value, much faster without any field-like torque (black solid line). The negative saturation gets delayed with increasing field-like torques within our simulated range.

## Section S10: Switching mechanism via micro-magnetic simulation

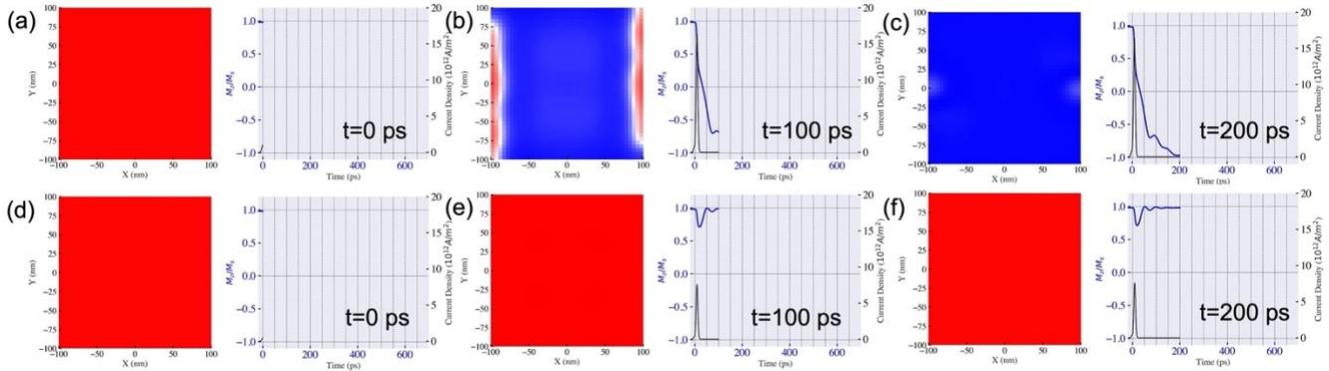

Fig. S10. (a-c) UBERMAG based micromagnetic simulation shows the coherent rotation of magnetization due to the application of the $9\ ps$ Gaussian current pulse with a current density of $16.5 \times 10^{12}\ A/m^2$ and (d-f) $7.6 \times 10^{12}\ A/m^2$ at three different time after the current pulse excitation. The color map shows the microscopic magnetization dynamics (red means $M_{+z}$ and blue means $M_{-z}$) at different timescale, the corresponding current profile and the normalized magnetization along z-axis is shown in the time-resolved graph.

We have used UBERMG[6,7] to microscopically simulate the magnetization dynamics using a simple model (at 0 K and without introducing any ultrafast heating-related demagnetization) using damping-like and field-like term in the LLG equation[8] and exchange, demagnetization and uniaxial anisotropy energy in calculating the total energy of the system. Without the ultrafast thermal anisotropy torque, we do not observe magnetization switching for a current density with $7.6 \times 10^{12}\ A/m^2$ as shown in Fig. S10 d-f. We need to increase the current density to $16.5 \times 10^{12}\ A/m^2$ to observe switching as shown in Fig. S10 a-c. This establishes the need for thermal anisotropy torque in observing the ultrafast switching with a smaller current density. We observe an ultrafast switching with zero crossing at $\sim 40\ ps$ and complete switching in $\sim 150\ ps$. We notice that a coherent rotation of the magnetization is the responsible mechanism for the switching phenomenon (even without the presence of any ultrafast demagnetization). We used a magnetic sample size of $100\ nm \times 100\ nm \times 1\ nm$ and the discretized cell size is $5\ nm \times 5\ nm \times 1\ nm$, hence basically it is a 2d model with no discretization along the z-axis. A $\sim 9\ ps$ Gaussian current pulse with a current density of $16.5 \times 10^{12}\ A/m^2$ and a $\Theta_{DL} = 0.2$ and $\Theta_{FL} = 0.04$; is used in the simulation where the other magnetic parameters (Saturation magnetization, Uniaxial anisotropy, and Damping Constant) are the same as used in the Macroscopic simulation model discussed in detail in Section 3 and the main manuscript. The current pulse is plotted along the right Y axis and the magnetization is plotted along the left Y axis. Magnetization of the sample turned from red ($M_{+z}$) to blue ($M_{-z}$) via a coherent rotation with time as a function of the current pulse propagation. We don't observe any formation of domain walls in the magnet as evident from the magnetization images. However, when using a $1\ ns$ current pulse (not shown here), we do observe the formation of the domain wall at the center of the sample and its propagation along the edges suggesting a domain-wall driven switching mechanism for $\sim ns$ long current pulse.